We summarize a few proposals for mixing F center states through the mediation of an appropriate symmetry-breaking vibrational mode. Electron-mode coupling energies odd-order in the mode coordinates are characteristic of the pseudo-Jahn-Teller mixing of nearly-degenerate opposite-parity electronic states mediated by an odd-parity vibrational mode. Coupling energies even-order in the vibrational coordinates lead to pseudo-Renner mixing of even-parity nearly-degenerate states or to dynamic-Renner mixing of degenerate states of whatever symmetry. Both Renner mixing energies add up to widen the crossover splitting of the adiabatic energies thereby enhancing the nonradiative deexcitation rate. The dynamic-Renner symmetry-breaking effects complement the dynamic Jahn-Teller mixing of degenerate electronic states by $1^{st}$-order coupling to an even-parity vibrational mode.

1. Introduction

One of the striking appearances in solid state physics is the breaking of configuration symmetry, due to the mixing of local electronic states by virtue of their coupling to symmetry-breaking vibrational modes. Textbook examples are provided by the Jahn-Teller effects which involve the mixing of degenerate electronic states or the pseudo-Jahn-Teller effects which involve the mixing of nearly-degenerate electronic states [1-4]. In symmetry terms these two appearances, vastly dissimilar for they comprise very different situations, have something in common which is their respective coupling energies being linear or $1^{st}$-order in the mode coordinate. We remind that the group theoretical requirement for the symmetry representation of the mixing vibration is that it be included in the direct product of the symmetry representations of the pair of electronic states.

In the quest for novel physical occurrences, solid state theorists have uncovered less traditional symmetry-breaking examples, due to the mixing of electronic states by their coupling to the higher-orders of the mode coordinates. One such example is the mixing produced by a coupling odd-order in the mode coordinates and a thrilling example is offered by the off-center ions in crystals, such as substitutional $Li^+$ in KCl. Indeed, while the off-center displacements themselves are $1^{st}$ order coupling in the $T_{1u}$-mode coordinates pushing the ion off-site along the body diagonals, the off-center ion's rotation around the central lattice site is $3^{rd}$ order in those same coordinates. The mixing electronic states responsible for the off-center displacements are identified as $a_{1g}$ and $t_{1u}$, respectively.

But there are other possibilities, yet to be proven viable in crystals. Among them is the Renner effect which has originally been advanced as a mixing of even-parity electron states by an even-parity vibration of atoms along a linear chain [1,3]. Unlike most of the previously mentioned examples, Renner's mixing energy is quadratic, not linear, in the mode coordinates. It should be added that the even-parity modes are the only ones allowed along a linear chain.

In what follows, we extend the definition of Renner's mixing energy so as to cover symmetry-breaking situations tractable by the even-order coupling terms. We shall see that the matter comprises not only what may be regarded an analogue to the pseudo-Jahn-Teller mixing, albeit of even parity electronic states, but also akin to dynamic Jahn-Teller mixing of degenerate electronic states of whatever symmetry. Renner's extensions may be found useful for situations where the the linear coupling terms are vanishing if they fall within the extremal range of the confugurational energy.

The present investigation is aimed at the multitude of electronic states and respective symmetries pertaining to the F center in alkali halides [5].

## 2. Pseudo-Jahn-Teller effect

### 2.1. Mixing constants

To begin with, we remind that the simplistic Pseudo-Jahn-Teller Effect (PJTE) is the mixing of two nearly-degenerate opposite-parity electronic states mediated by an odd-parity vibration. This definition suggests just how it can be extended so as to cover pairs of more involved symmetries. Considering the particular constitution of the $T_{1u}$ vibrational mode of the six nn cations around the anion vacancy, these form three cation pairs each performing in-phase displacements along the three respective crystallographic directions. See Figure 1 for an illustration of the pseudo-Jahn-Teller effect considered applicable to the $T_{1u}$ mixing of 2s- and 2p-like F center states.

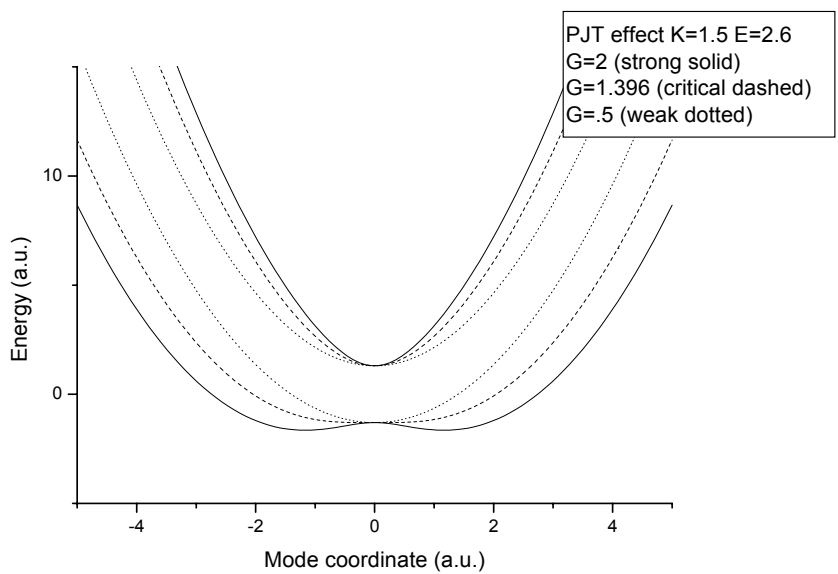

Figure 1

Vibronic potential energies (upper & lower branches) of the linear pseudo-Jahn-Teller effect. Three pairs of potentials are illustrated corresponding to 3 numerical values of the electron-vibrational mode coupling parameter G according to the legend in insert. The situation applies to the mixing of 2p- & 2s-like F center states by the $T_{1u}$ mode. The depicted potentials follow the equation $E_{\pm}(q)_{PJTE} = \frac{1}{2}Kq^2 \pm \frac{1}{2}\sqrt{(4G^2q^2+E_{12}^2)}$ where q is a 1D mode coordinate, K is the stiffness, G is the coupling constant, and $E_{12}$ is the interlevel energy gap [1-4].

For example, displacements along <100>, <010>, and <001> will result in a deformation of the cavity along <111>, etc. The cavity distortions give rise to an electrostatic potential in the point-ion field modulated by the displacements $q_k$:

$$U(r_0,q_l) = \alpha_M e / [(x_0 + q_x)^2 + (y_0 + q_y)^2 + (z_0 + q_z)^2]^{1/2}. \qquad (1)$$

$\alpha_M$, $V_M = \alpha_M e^2/r_0$, and $r_0 = (x_0^2+y_0^2+z_0^2)^{1/2}$ are Madelung's constant, potential, and 'cavity radius'. $U(r_0,q_l)$ generates an electric field $F(r_0,q_l) = -\text{grad}_q U(r_0,q_l)$ which couples to the electric dipole $\mathbf{p} = e\mathbf{r}$ mixing the electronic states involved. The coupling energy $V = -e\mathbf{r}\cdot\mathbf{F}$ is

$$V(r_0,q_l) = -(\alpha_M e)\{ex(x_0+q_x)+ey(y_0+q_y)+ez(z_0+q_z)\}$$

$$/ [(x_0+q_x)^2+(y_0+q_y)^2+(z_0+q_z)^2]^{3/2} \qquad (2)$$

To derive electron-phonon coupling operators of some order, we differentiate $V(r_0,q_l)$ in $q_i$ to that order. Details are to be found in Appendix I. Mixing constants are calculated as off-diagonal matrix elements of the coupling operators in basis states which are eigenstates of the semi-continuum electronic potential, i.e. spherical Bessel functions inside the cavity and hydrogen-like wave functions outside the cavity:

$u_{ij...k,xyz} = \langle 2t_{1u,xyz} | u_{ij...k}(r) | 1a_{1g}\rangle$

$= A_{\kappa 1}A_{\kappa'0} \int u_{ij...k}(r)\{[sin(\kappa r)/(\kappa r)-cos(\kappa r)]/(\kappa r)\}\{sin(\kappa' r)/(\kappa' r)\}r^2 sin\theta dr d\theta d\varphi \big|_{r\leq r_0}$

$+ B_{21}B_{10} \int u_{ij...k}(r)\, \mathbf{r}_{xyz}\, exp[-(\alpha_{2t1u}+\alpha_{1a1g})r]r^2 sin\theta dr d\theta d\varphi \big|_{r\geq r_0}$

$= A_{\kappa 1}A_{\kappa'0}\,_0\!\int^{r_0} dr\,_0\!\int^{2\pi} d\varphi\,_0\!\int^{\pi} d\theta\, u_{ij...k}(r)\{[sin(\kappa r)/(\kappa r)-cos(\kappa r)]/(\kappa r)\}$

$\times \{sin(\kappa' r)/(\kappa' r)\}r^2 sin\theta$

$+ B_{21}B_{10}\,_{r0}\!\int^{0} dr\,_0\!\int^{2\pi} d\varphi\,_0\!\int^{\pi} d\theta\, u_{ij...k}(r)\, \mathbf{r}_{xyz}\, r^2 sin\theta\, exp[-(\alpha_{2t1u}+\alpha_{1a1g})r],$

$v_{ij...k,xyz} = \langle 2t_{1u,xyz} | v_{ij...k}(r) | 2a_{1g}\rangle$

$= A_{\kappa 1}A_{\kappa'0} \int v_{ij...k}(r)\{[sin(\kappa r)/(\kappa r)-cos(\kappa r)]/(\kappa r)\}\{sin(\kappa' r)/(\kappa' r)\}r^2 sin\theta dr d\theta d\varphi \big|_{r\leq r_0}$

$+ B_{21}B_{20} \int v_{ij...k}(r)\, \mathbf{r}_{xyz}\, 2(1-\alpha_{2a1g}r)exp[-(\alpha_{2t1u}+\alpha_{2a1g})r]r^2 sin\theta dr d\theta d\varphi \big|_{r\geq r_0}$

$$= A_{\kappa l} A_{\kappa' 0} \int_0^{r_0} dr \int_0^{2\pi} d\varphi \int_0^{\pi} d\theta \, v_{ij\ldots k} I\{[sin(\kappa r)/(\kappa r) - cos(\kappa r)]/(\kappa r)\}$$

$$\times \{sin(\kappa' r)/(\kappa' r)\} r^2 sin\theta$$

$$+ B_{21} B_{20} \int_{r_0}^{\infty} dr \int_0^{2\pi} d\varphi \int_0^{\pi} d\theta \, v_{ij\ldots k}(r) \, \mathbf{r}_{xyz} 2(1-\alpha_{2a1g} r) r^2 sin\theta \, exp[-(\alpha_{2t1u}+\alpha_{2a1g})r],$$

where $u_{ij\ldots k}$ and $v_{ij\ldots k}$ are derivatives of $V(r_0,q_l)$ in $q_i$, to be derived analytically or calculated numerically. Here $A_{\kappa l}$, $B_{nl}$ are the normalization constants, and $\kappa = u / r_0$, are determined by the continuity conditions for the wavefunctions at $r = r_0$ [5]. The following mixing constants are derived:

(i) in-cavity for $r \leq r_0$:

$$b_{x,x} = -(8\pi/3)(V_M/r_0^2)P(v,v'),$$

$$d_{xxx,x} = -(24\pi/5)(V_M/r_0 4)P(v,v'),$$

$$d_{xxz,z} = -4(1+22\pi/5)(V_M/r_0^4)P(v,v').$$

(ii) out-of-cavity for $r \geq r_0$:

$$b_{x,x} = -(8\pi/3)(V_M/r_0) N_{a1g} N_{t1u}(1/\alpha^4)$$

$$\times [exp(-\alpha r_0)/(\alpha r_0)][(\alpha r_0)^4 + \sum_{k=1}^{4} 4 \times 3 \times \ldots \times (4-k+1)(\alpha r_0)^{4-k}],$$

$$d_{xxx,x} = -(12/5)\pi(V_M/r_0^3) N_{a1g} N_{t1u} (1/a^4)$$

$$\times [exp(-\alpha r_0)/(\alpha r_0)][(\alpha r_0)^4 + \sum_{k=1}^{4} 4 \times 3 \times \ldots \times (4-k+1)(\alpha r_0)^{4-k}],$$

$$d_{xxz,z} = -6\pi(64/15)(V_M/r_0^3) N_{a1g} N_{t1u}(1/\alpha^4)$$

$$\times [exp(-\alpha r_0)/(\alpha r_0)][(\alpha r_0)^4 + \sum_{k=1}^{4} 4 \times 3 \times \ldots \times (4-k+1)(\alpha r_0)^{4-k}].$$

Here $\alpha_{1s2p} = \alpha_{2t1u}+\alpha_{1a1g} = (3/2)(Z/\alpha_0)$ or $\alpha_{2s2p}= \alpha_{2t1u}+\alpha_{2a1g}=(Z/\alpha_0)$ giving $N_{a1g}N_{t1u} (1/\alpha^4) = (2/3)^4 (1/4\pi\sqrt{2})$ for 1s-2p and $N_{a1g}N_{t1u}(1/\alpha^4) = (1/4\pi\sqrt{2})$ for 2s-2p. The semi-continuum hydrogen-like wave function normalization constants $B_{nl}$ of Table I relate to $N_{1a1g}$ by way of:

$$N_{1a1g} = \pi^{-\frac{1}{2}}(Z/a_0)^{3/2} = \pi^{-\frac{1}{2}}\alpha_{1a1g}^{3/2} = (1/\sqrt{\pi})u_{1a1g}^{3/2} r_0^{-3/2} = (1/4\pi)^{\frac{1}{2}} B_{10}$$

$$N_{2t1u} = \pi^{-\frac{1}{2}}(Z/2\alpha_0)^{5/2} = \pi^{-\frac{1}{2}}\alpha_{2t1u}^{5/2} = (1/\sqrt{\pi})\alpha_{2t1u} u_{2t1u}^{3/2} r_0^{-3/2} = (3/4\pi)^{\frac{1}{2}}\alpha_{2t1u} B_{21}$$

$$N_{2a1g} = \pi^{-\frac{1}{2}}(Z/2\alpha_0)^{3/2} = \pi^{-\frac{1}{2}}\alpha_{2a1g}^{3/2} = (1/\sqrt{\pi})u_{2a1g}^{3/2} r_0^{-3/2} = (1/4\pi)^{\frac{1}{2}} B_{20}$$

(iii) Complete mixing constants:

The complete mixing constant tensors obtain by summing up in-cavity and out-of-cavity contributions weighted by the respective wave function fractions:

$$G_{1,ns2p} = -(8\pi/3)(V_M/r_0)\{[P(v,v')/r_0] + (\sqrt{3}/4\pi)\alpha_{2p}B_{10}B_{21}(1/\alpha)^4$$

$$\times [exp(-\alpha r_0)/(\alpha r_0)][(\alpha r_0)^4 + \sum_{k=1}^{4} 4\times 3\times...\times(4-k+1)(\alpha r_0)^{4-k}]\},$$

$$T_{111,ns2p} = -(12/5)\pi(V_M/r_0^3)\{2[P(v,v')/r_0] + (\sqrt{3}/4\pi)\alpha_{2p}B_{10}B_{21}(1/\alpha)^4$$

$$\times [exp(-\alpha r_0)/(\alpha r_0)][(\alpha r_0)^4 + \sum_{k=1}^{4} 4\times 3\times...\times(4-k+1)(\alpha r_0)^{4-k}]\},$$

$$T_{113,ns2p} = -(4/5)(V_M/r_0^3)\{(5+22\pi)[P(v,v')/r_0] + 8\sqrt{3}\alpha_{2p}B_{10}B_{21}(1/\alpha)^4$$

$$\times [exp(-\alpha r_0)/(\alpha r_0)][(\alpha r_0)^4 + \sum_{k=1}^{4} 4\times 3\times...\times(4-k+1)(\alpha r_0)^{4-k}]\}.$$

Herein $\alpha = \alpha_{2p} + \alpha_{ns} \equiv \alpha_{2t1u} + \alpha_{na1g}$ and $P(v,v')$ are polynomials [5].

### 3. Renner effects

#### 3.1. Pseudo-Renner effect in 1s-2s mixing

With regard to the $T_{1u}$ odd-mode coupling to F center states, we distinguish between odd-order mixing ('pseudo-Jahn-Teller effect') and even-order mixing. We refer to the latter as 'pseudo-Renner effect' (PRE). In so far as the pseudo-Jahn-Teller effect has been the subject of a number of detailed studies relevant to the F center, we will next illustrate the pseudo-Renner mixing, say of the 1s- and 2s-like eigenstates of the semi-continuum electronic potential. From the 1$^{st}$ square of the secular determinant we get the following equation for the adiabatic eigenvalues:

$$(E - \tfrac{1}{2}Kq^2 - E_{1s})(E - \tfrac{1}{2}Kq^2 - E_{2s}) - (C_{12}q^2 + D_{12})(C_{12}q^2 + D_{12}) \equiv$$

$$E^2 - [(\tfrac{1}{2}Kq^2 - E_{1s}) + (\tfrac{1}{2}Kq^2 - E_{2s})]E + (\tfrac{1}{2}Kq^2 - E_{1s})(\tfrac{1}{2}Kq^2 - E_{2s}) - (C_{12}q^2 + D_{12})(C_{12}q^2 + D_{12}) = 0$$

The energy eigenvalues are:

$$E_{\pm}(q) = \tfrac{1}{2}[(\tfrac{1}{2}Kq^2 - E_{1s}) + (\tfrac{1}{2}Kq^2 - E_{2s})] \pm \tfrac{1}{2}\sqrt{\{[(\tfrac{1}{2}Kq^2 - E_{1s}) + (\tfrac{1}{2}Kq^2 - E_{2s})]^2}$$

$$- 4[(\tfrac{1}{2}Kq^2 - E_{1s})(\tfrac{1}{2}Kq^2 - E_{2s}) - (C_{12}q^2 + D_{12})(C_{12}q^2 + D_{12})]\}$$

$$= \tfrac{1}{2}Kq^2 - \tfrac{1}{2}(E_{1s} + E_{2s}) \pm \tfrac{1}{2}\sqrt{\{4(C_{12}q^2 + D_{12})^2 + (E_{2s} - E_{1s})^2\}}$$

$$= \tfrac{1}{2}Kq^2 - \tfrac{1}{2}(E_{1s} + E_{2s}) \pm \tfrac{1}{2}\sqrt{\{4C_{12}^2 q^4 + (E_{2s} - E_{1s})^2\}} \quad (D_{12} \equiv 0)$$

Differentiating with respect to q,

$$dE_{\pm}(q)/dq = Kq \pm 8C_{12}^2 q^3/\sqrt{\{4C_{12}^2 q^4 + (E_{2s} - E_{1s})^2\}}$$

we see a central extremum appearing at q=0. At that point, the two adiabatic branches split, the gap amounting to $E_{2s} - E_{1s}$ as in the pseudo-Jahn-Teller effect. There are also

two side extrema whose positions are found by rationalizing and solving for q to be at:

$$\pm q_0 = \pm[K|E_{2s}-E_{1s}| / 2C_{12}\sqrt{(16C_{12}^2-K^2)}]^{1/2}$$

$q_0$ is real for $\beta = 4C_{12} / K > 1$, the condition for a strong pseudo-Renner-effect mixing. No side extrema appear at $\beta = 4C_{12} / K < 1$ which is the weak-mixing condition. At $\beta = 4C_{12} / K = 1$ the side extrema are asymptotic. The second derivative gives the curvature at q:

$$d^2E_\pm(q)/dq^2 = K \pm 8C_{12}^2 q^2 [\{4C_{12}^2 q^4 + 3(E_{2s}-E_{1s})^2\}] / \{4C_{12}^2 q^4 + (E_{2s}-E_{1s})^2\}^{3/2}$$

At q = 0, the curvature amounts to K > 0 (minimum), while the lateral curvatures are

$$K_{0\pm} = K\{1 \pm 4C_{12}[K^2/(16C_{12}^2-K^2)+3] / \sqrt{(16C_{12}^2-K^2)[K^2/(16C_{12}^2-K^2)+1]^3}\} \equiv$$

$$K\{1 \pm \beta[1/(\beta^2-1)+3]/[(\beta^2-1)[1/(\beta^2-1)+1]^3]^{1/2}\}$$

at $\pm q_0$. $K_{0\pm}$ is seen to be dependent on $\beta = 4C_{12} / K$, though not on $E_{2s}-E_{1s}$.

It might also be instructive to calculate the energy difference on the lower adiabatic energy branch between the $q = \pm q_0$ and $q = 0$ points: $E_D = E_-(\pm q_0) - E_-(0)$. We have

$$E_-(0) = -\tfrac{1}{2}(E_{1s}+E_{2s}) - \tfrac{1}{2}(E_{2s}-E_{1s}) = -E_{2s}$$

and using $q_0$ we get

$$E_-(\pm q_0) = (E_{2s}-E_{1s})[1/\sqrt{(\beta^2-1)}](1/\beta - \tfrac{1}{2}\beta) - \tfrac{1}{2}(E_{1s}+E_{2s})$$

Therefore, the energy difference reads

$$E_D = (E_{2s}-E_{1s})\{[1/\sqrt{(\beta^2-1)}](1/\beta - \tfrac{1}{2}\beta) + \tfrac{1}{2}\}$$

This difference vanishes, $E_D \sim 0$, at large $\beta \gg 1$. At $\beta \sim 1$, $\beta = 1+\Delta\beta$ ($\Delta\beta \ll 1$):

$$E_D \sim |E_{2s}-E_{1s}|[1/2\sqrt{(2\Delta\beta)}]$$

$E_D \to \infty$ at $\beta \to 1$ ($\Delta\beta \to 0$).

We see that $\beta = 4C_{12}/K$ controls the configurational energy as does $4E_{JT}/(E_{2s}-E_{1s})$ in the linear Jahn-Teller theory. Accordingly, there are three parameter ranges: weak coupling ($\beta < 1$), critical coupling ($\beta = 1$), and strong coupling ($\beta > 1$). The asymptotic behaviour of Renner's vibronic potential energy is $E_\pm(q) \propto \tfrac{1}{2}(K\pm 2C_{12})q^2 = \tfrac{1}{2}K(1\pm \tfrac{1}{2}\beta)q^2$ at $q \to \infty$. Vibronic potentials for the pseudo-Renner effect are shown in Figure 2.

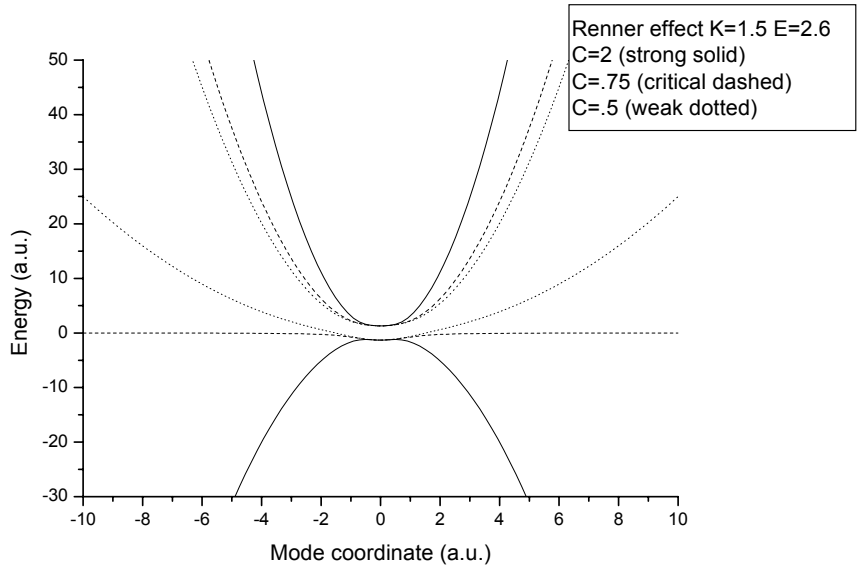

Figure 2

Vibronic potential energies (upper & lower branches) of the quadratic pseudo-Renner effect. Three pairs of potentials are illustrated corresponding to 3 numerical values of the coupling constant (β = 2$C_{12}$/K = 2.67, 1 and 0.67) according to the legend in insert. The situation may apply to the $A_{1g}$ mixing of 1s- & 2s-like F center states. The depicted potentials follow the equation $E_\pm(q)_{PRE} = \frac{1}{2}Kq^2 \pm \frac{1}{2}\sqrt{4C_{12}^2 q^4 + E_{12}^2}$ where q is the 1D mode coordinate, K is the stiffness, C ≡ $C_{12}$ is the coupling constant, and $E_{12}$ is the interlevel energy gap.

### 3.2. Dynamic Renner effect in $2p_x$-$2p_y$ mixing

There being a vast amount of literature on the dynamic Jahn-Teller effect [1-4], we will rather focus on the similarities and differences with its 2$^{nd}$ power analogues. Most of them will be identified by the attentive readers themselves.

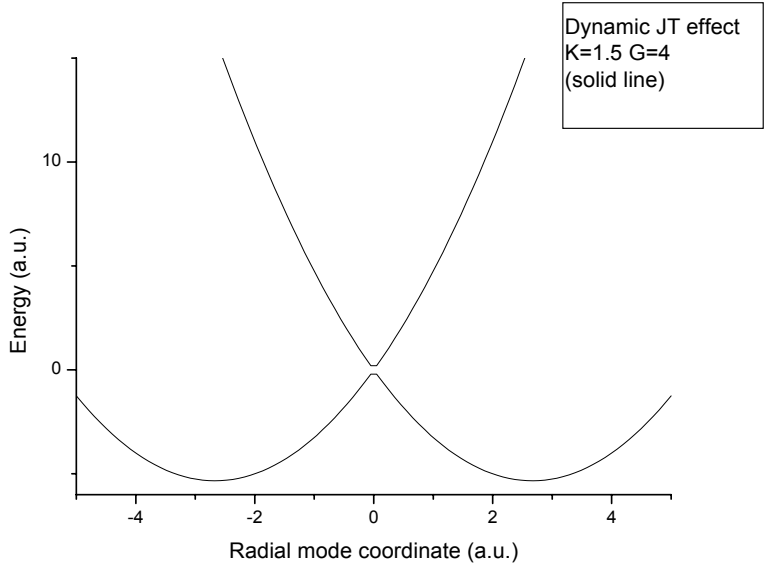

Figure 3

Vibronic potential energies (upper and lower branches) for the linear dynamic Jahn-Teller effect in the vibronic mixing of degenerate $2p_i$-like F center states (i = x,y,z). The depicted potentials follow the equation $E_\pm(\mathbf{q})_{JTE} = \tfrac{1}{2}\sum_i K_i q_i^2 \pm \sqrt{\{(2G_x q_x)^2 + (2G_y q_y)^2 + (2G_z q_z)^2\}}$ of two displaced harmonic parabolas where $q_x$, etc. are mode coordinates, K is the stiffness and $G_x$, etc. are the coupling constants. A small dynamic energy splitting at crossover makes transitions between the lateral wells possible at all.

In the original higher symmetry configuration, the 2p-like excited state is triply degenerate, such that the pseudo-Renner mechanism does not apply. For degenerate electronic states an extension of the dynamic vibronic effect is conceivable. We again select a secular determinant, now 3$^{rd}$ order:

$(E-\tfrac{1}{2}\sum_i K_i q_i^2 - E_{2p})^3 +$

$(C_{yx} q_y q_x + D_{yx})(C_{zy} q_z q_y + D_{zy})(C_{xz} q_x q_z + D_{xz}) +$

$(C_{xy} q_x q_y + D_{xy})(C_{yz} q_y q_z + D_{yz})(C_{zx} q_z q_x + D_{zx}) -$

$(C_{xz} q_x q_z + D_{xz})(E-\tfrac{1}{2}\sum_i K_i q_i^2 - E_{2p})(C_{zx} q_z q_x + D_{zx}) -$

$(C_{xy} q_x q_y + D_{xy})(C_{yx} q_y q_x + D_{yx})(E-\tfrac{1}{2}\sum_i K_i q_i^2 - E_{2p}) -$

$(C_{yz} q_y q_z + D_{yz})(C_{zy} q_z q_y + D_{zy})(E-\tfrac{1}{2}\sum_i K_i q_i^2 - E_{2p}) \equiv$

$(E-\tfrac{1}{2}\sum_i K_i q_i^2 - E_{2p})\{(E-\tfrac{1}{2}\sum_i K_i q_i^2 - E_{2p})^2 - (C_{xz} q_x q_z + D_{xz})(C_{zx} q_z q_x + D_{zx}) -$

$(C_{xy}q_xq_y+D_{xy})(C_{yx}q_yq_x+D_{yx}) - (C_{yz}q_yq_z+D_{yz})(C_{zy}q_zq_y+D_{zy})\} +$

$(C_{yx}q_yq_x+D_{yx})(C_{zy}q_zq_y+D_{zy})(C_{xz}q_xq_z+D_{xz}) +$

$(C_{xy}q_xq_y+D_{xy})(C_{yz}q_yq_z+D_{yz})(C_{zx}q_zq_x+D_{zx}).$

The last two terms above are small in so far as they contain contrubitions $6^{th}$ order in $q_i$ at $D_{ij} = 0$. If these free terms are neglected, then the secular equation acquires the roots:

$E_1 = \tfrac{1}{2}\sum_i K_i q_i^2 + E_{2p}$

$E_{2,3} = \tfrac{1}{2}\sum_i K_i q_i^2 + E_{2p} \pm \{(C_{xz}q_xq_z+D_{xz})(C_{zx}q_zq_x+D_{zx}) +$

$(C_{xy}q_xq_y+D_{xy})(C_{yx}q_yq_x+D_{yx}) + (C_{yz}q_yq_z+D_{yz})(C_{zy}q_zq_y+D_{zy})\}^{\frac{1}{2}} \equiv$

$\tfrac{1}{2}\sum_i K_i q_i^2 + E_{2p} \pm \{(C_{xz}q_xq_z+D_{xz})^2 + (C_{xy}q_xq_y+D_{xy})^2 + (C_{yz}q_yq_z+D_{yz})^2\}^{\frac{1}{2}},$

the latter form accounting for the symmetry with respect to interchanging the coordinate axes.

The complete equation is solved in a somewhat complicated form. Following the familiar algebra, a real root of the cubic equation $x^3+px+q = 0$ is given by Cardano's formula:

$x_1 \sim E_1 - \tfrac{1}{2}\sum_i K_i q_i^2 - E_{2p} = \{-(q/2)+[(q/2)^2+(p/3)^3]^{1/2}\}^{1/3}$

$+ \{-(q/2)-[(q/2)^2+(p/3)^3]^{1/2}\}^{1/3}$

with

$p = -\{(C_{xz}q_xq_z+D_{xz})(C_{zx}q_zq_x+D_{zx}) + (C_{xy}q_xq_y+D_{xy})(C_{yx}q_yq_x+D_{yx})$

$+ (C_{yz}q_yq_z+D_{yz})(C_{zy}q_zq_y+D_{zy})\}$

$q = (C_{yx}q_yq_x+D_{yx})(C_{zy}q_zq_y+D_{zy})(C_{xz}q_xq_z+D_{xz})$

$+ (C_{xy}q_xq_y+D_{xy})(C_{yz}q_yq_z+D_{yz})(C_{zx}q_zq_x+D_{zx})$

Once that root is at hand the remaining two roots are found by solving the quadratic equation $x^2+x_1x+(p+x_1^2) \equiv x^2+x_1x - (q/x_1) = 0$, as obtained from $(x-x_1)(x^2+x_1x-q/x_1) \equiv x^3+px+q$:

$x_{2,3} \sim E_{2,3} - \tfrac{1}{2}\sum_i K_i q_i^2 - E_{2p} = -\tfrac{1}{2}x_1 \pm \tfrac{1}{2}[x_1^2 - 4(p+x_1^2)]^{\frac{1}{2}} = -\tfrac{1}{2}x_1 \pm \tfrac{1}{2}[x_1^2+4(q/x_1)]^{\frac{1}{2}}.$

At $q^2/p^3 \ll 1$ we get

$x_1 \equiv E_1 - \tfrac{1}{2}\sum_i K_i q_i^2 - E_{2p} \sim$

$\sqrt{(p/3)}\{1-(1/3)(q/2)/(p/3)^{3/2}\}-\sqrt{(p/3)}\{1+(1/3)(q/2)/(p/3)^{3/2}\}= -q/p$

$x_{2,3} \equiv E_{2,3} - \frac{1}{2}\sum_i K_i q_i^2 - E_{2p} \sim \frac{1}{2}(q/p) \pm \frac{1}{2}\{(q/p)^2 - 4[p+(q/p)^2]\}^{\frac{1}{2}} \sim \frac{1}{2}(q/p) \pm \frac{1}{2}[(q/p)^2 - 4p]^{\frac{1}{2}}$

which yields $x_{2,3} \sim \pm\sqrt{(-p)}$ at $q = 0$, as above.

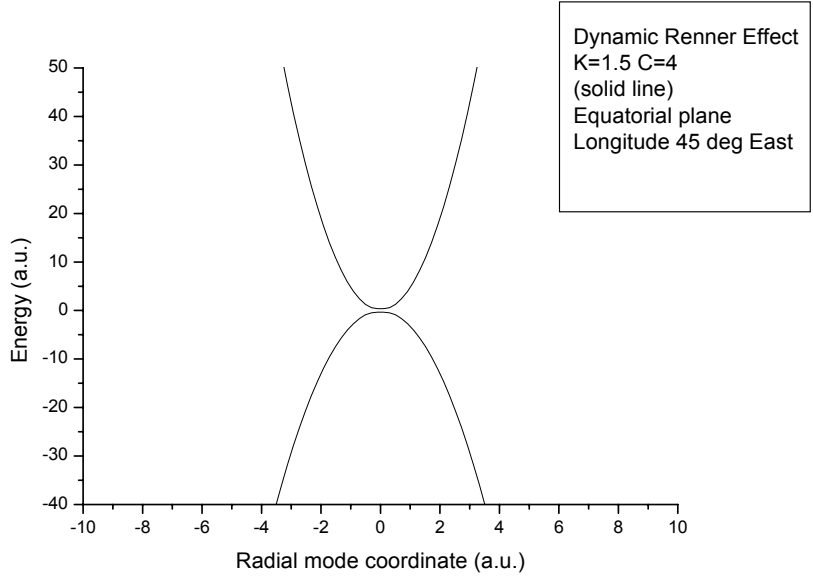

Figure 4

Vibronic potential energies (upper and lower branches) for the quadratic dynamic Renner effect in the vibronic mixing of degenerate $2p_i$-like F center states (i = x,y,z). The depicted potentials follow the equation $E_\pm(\mathbf{q})_{RE} = \frac{1}{2}\sum_i K_i q_i^2 \pm \sqrt{\{(2C_{xz}q_x q_z)^2 + (2C_{xy}q_x q_y)^2 + (2C_{yz}q_y q_z)^2\}}$ of two displaced 4$^{th}$ power parabolas, where $q_x$, etc. are mode coordinates, K is the stiffness and $C_{xy}$, etc. are the coupling constants. The dynamic crossover energy splitting enables transitions between the lateral wells.

One should note that the dynamic-Renner effect splits the adiabatic potential energy surface into lower and upper branches, the splitting term being second-order in the vibrational coordinates.

### 3.3. Mixing constants

States of the same parity may only mix through the even-order coupling terms in the interaction expansion. Concomitantly we define 2$^{nd}$-order mixing constants between same-parity states, even-even and odd-odd, namely:

$c_{1s2s} = <1s | \partial^2 V(r_0,q_l)/\partial q_x^2 |_{q=0} | 2s> \sim <1s | 3(rr_0 + 3xx_0)/r_0^5 | 2s>$

$c_{xy} = <2p_x | \partial^2 V(r_0,q_l)/\partial q_x \partial q_y |_{q=0} | 2p_y> = <2p_x | 9xy_0/r_0^5 | 2p_y>$,

etc. These are computed using semicontinuum wavefunctions, as above. Details can be found in Appendix II. The former constant is

$c_{1s2s} = \alpha_M e^2 \{(12\pi/vv')A_{\kappa 0}A_{\kappa' 0}\{[(v-v')^{-1}\sin(v-v')-(v+v')^{-1}\sin(v+v')]$

$+ [(v-v')^{-2}\cos(v-v')-(v+v')^{-2}\cos(v+v')]\}$

$+ B_{10}B_{20}48\pi[\exp(-[u_{1s}+u_{2s}])/(u_{1s}+u_{2s})^4] \times$

$\{[u^3 + \sum_{k=1}^{3} 3 \times 2 \times ... \times (3-k)u^{3-k}] - [u_{2s}/(u_{1s}+u_{2s})][u^4 + \sum_{k=1}^{4} 4 \times 3 \times ... \times (4-k)u^{4-k}]\}\}$

with $\alpha = \alpha_{1s} + \alpha_{2s}$, $u = \alpha r_0 = u_{1s} + u_{2s}$. The latter constant is

$c_{xy} = \alpha_M e^2 \{(6\pi/5)A_{\kappa 1}A_{\kappa' 1}\{\{[(v-v')^{-1}\sin(v-v')-(v+v')^{-1}\sin(v+v')]$

$+ [(v-v')^{-2}\sin(v-v')-(v+v')^{-2}\sin(v+v')]\}/vv'$

$+ (1/v + 1/v')\sum_{k=0}^{2} k!(2k)\{(v+v')^{-(k+1)}\cos(v+v'+\frac{1}{2}k\pi) + (v-v')^{-(k+1)}\cos(v-v'+\frac{1}{2}k\pi)\}$

$+ \sum_{k=0}^{3} k!(3k)\{(v+v')^{-(k+1)}\sin(v+v'+\frac{1}{2}k\pi) + (v-v')^{-(k+1)}\sin(v-v'+\frac{1}{2}k\pi)\}\}$

$+ (12/5)\pi B_{21}B_{21} [\exp(-u)/u^4][u^3 + \sum_{k=1}^{3} 3 \times 2 \times ... \times (3-k)u^{3-k}]\}$

at $v \neq v'$. Alternatively, when $v = v'$:

$c_{xy} = \alpha_M e^2 \{(6\pi/5)A_{\kappa 1}A_{\kappa' 1} \times \{\frac{1}{2}/v^2 - [(2v)^{-1}\sin(2v)+(2v)^{-2}\cos(2v)]/v^2$

$- (2/v)\sum_{k=0}^{2} k!(2k)(2v)^{-(k+1)}\cos(2v+\frac{1}{2}k\pi) + \sum_{k=0}^{3} k!(3k)(2v)^{-(k+1)}\sin(2v+\frac{1}{2}k\pi)\}$

$+ (12/5)\pi B_{21}B [\exp(-u)/u^4][u^3 + \sum_{k=1}^{3} 3 \times 2 \times ... \times (3-k)u^{3-k}]\}$

with $\alpha = 2\alpha_{2p}$, $u = \alpha r_0 = 2\alpha_{2p}r_0 = 2u_{2p}$.

There we have the constants for a pseudo-Renner mixing of the first three semicontinuum eigenstates of the F center in an alkali halide. Calculated numerical values are listed in Table II.

## 4. Conclusion

We considered electronic states of various symmetries at the F center in alkali halides. Our intimate goal was unravelling the mixing mechanisms of these electronic states in looking for ways to determine the realistic electronic crossover gap so as to estimate the efficiency of the nonradiative deexcitation rate. Among the options taken in view were the linear-coupling Jahn-Teller effects and the quadratic-coupling Renner effects. While the former involved even-even and odd-even pair of states mixed by even-parity modes (JTE) or odd-parity modes (PJTE), respectively, the latter required even-even (PRE) or odd-odd (DRE) pairs mixed by even-parity modes. Mixing constants at the center of the debate are derived for all the cases under consideration.

All these applied to the coupling of ground and excited F center states. The present paper theorized on the conceivable mechanism though the particular examples would be dealt with in a subsequent Part III.

## Appendix I

### Deriving the $T_{1u}$-coupling constants
#### AI.1. Definition

The electron-mode coupling energy being

$$V(r_0,q_l) = -(\alpha_M e)\{ex(x_0+q_x) + ey(y_0+q_y) + ez(z_0+q_z)\}/[(x_0+q_x)^2+(y_0+q_y)^2+(z_0+q_z)^2]^{3/2}$$

we differentiate it in $q_i$ to derive mixing operators. Introducing spherical coordinates

$x/r = x_0/r_0 = cos\varphi sin\theta$

$y/r = y_0/r_0 = sin\varphi sin\theta$

$z/r = z_0/r_0 = cos\theta$

and dropping the $\alpha_M e^2$ factor for simplicity, we get:

$V(r_0,q_l)|_{q=0} = r/r_0^2$

$\partial V(r_0,q_l)/\partial q_x|_{q=0} = -2x/r_0^3$

$\partial^2 V(r_0,q_l)/\partial q_x \partial q_y|_{q=0} = 9xy_0/r_0^5$

$\partial^2 V(r_0,q_l)/\partial q_x^2|_{q=0} = 3(rr_0+3xx_0)/r_0^5$

$\partial^3 V(r_0,q_l)/\partial q_x \partial q_y \partial q_z|_{q=0} = 74xy_0z_0/r_0^7$

$\partial^3 V(r_0,q_l)/\partial q_x^2 \partial q_z|_{q=0} = -6z_0(rr_0+11xx_0)/r_0^7$

$\partial^3 V(r_0,q_l)/\partial q_x|_{q=0} = 6x(3r_0^2-11x_0^2)/r_0^7$

The mixing constants of different order in the electron-mode coupling are:

$u_{ij..k,xyz} = <2t_{1u,xyz} | u_{ij...k}(\mathbf{r}) | 1a_{1g}>$

$\sim \int \mathbf{r}_{xyz} u_{ij...k}(\mathbf{r}) exp[-(\alpha_{2t1u}+\alpha_{1a1g})r] r^2 sin\theta dr d\theta d\varphi$

$= {}_0\!\int^\infty dr \, {}_0\!\int^{2\pi} d\varphi \, {}_0\!\int^\pi d\theta \, \mathbf{r}_{xyz} u_{ij...k}(r) r^2 sin\theta exp[-(\alpha_{2t1u}+\alpha_{1a1g})r]$

$v_{ij...k,xyz} = <2t_{1u,xyz} | u_{ij...k}(\mathbf{r}) | 2a_{1g}>$

$\sim \int \mathbf{r}_{xyz} v_{ij...k}(\mathbf{r}) 2(1-\alpha_{2a1g}r) exp[-(\alpha_{2t1u}+\alpha_{2a1g})r] r^2 sin\theta dr d\theta d\varphi$

$= {}_0\!\int^\infty dr \, {}_0\!\int^{2\pi} d\varphi \, {}_0\!\int^\pi d\theta \, \mathbf{r}_{xyz} u_{ij...k}(r) 2(1-\alpha_{2a1g}r) r^2 sin\theta exp[-(\alpha_{2t1u}+\alpha_{2a1g})r]$

where $u_{ij...k}(r)$ are the mixing operators, derivatives of the coupling energy in $q_i$.

For estimating the coupling constants, we use semi-continuum potential wave functions $\psi(r) = j_l(\kappa r)$ for $r \leq r_0$ and $\psi(r) = R_{nl}(\alpha r)$ for $r \geq r_0$, where $r_0$ is the cavity radius; $j_l(\kappa r)$ are the spherical Bessel functions and $R_{nl}(\alpha r)$ are the hydrogen-like wave functions. The radial in-cavity states are, respectively:

$|1a_{1g}> \sim sin(\kappa r)/(\kappa r)$

$|2t_{1u}> \sim [sin(\kappa r)/(\kappa r) - cos(\kappa r)]/(\kappa r)$

$|2a_{1g}> \sim sin(\kappa r)/(\kappa r)$.

The out-of-cavity normalized hydrogen-like wavefunctions are:

$|1a_{1g}> = \pi^{-1/2}(Z/a_0)^{3/2} exp(-Zr/a_0) = N_{1a1g} exp(-\alpha_{1a1g}r)$

$|2t_{1u,x}> = (32\pi)^{-1/2}(Z/a_0)^{3/2}(Zr/a_0) exp(-Zr/2a_0) cos\varphi sin\theta = N_{2t1u} x \, exp(-\alpha_{2t1u}r)$

$|2t_{1u,y}> = (32\pi)^{-1/2}(Z/a_0)^{3/2}(Zr/a_0) exp(-Zr/2a_0) sin\varphi sin\theta = N_{2t1u} y \, exp(-\alpha_{2t1u}r)$

$|2t_{1u,z}> = (32\pi)^{-1/2}(Z/a_0)^{3/2}(Zr/a_0) exp(-Zr/2a_0) cos\theta = N_{2t1u} z \, exp(-\alpha_{2t1u}r)$

$|2a_{1g}> = \pi^{-1/2}(Z/2a_0)^{3/2} 2(1-Zr/2a_0) exp(-Zr/2a_0) = N_{2a1g} 2(1-\alpha_{2a1g}r) exp(-\alpha_{2a1g}r)$,

$N_{1a1g} = \pi^{-1/2}(Z/a_0)^{3/2} = \pi^{-1/2}\alpha_{1a1g}^{3/2}$, $\alpha_{1a1g} = Z/a_0$

$N_{2t1u} = \pi^{-1/2}(Z/2a_0)^{5/2} = \pi^{-1/2}\alpha_{2t1u}^{5/2}$, $\alpha_{2t1u} = Z/2a_0$

$N_{2a1g} = \pi^{-1/2}(Z/2a_0)^{3/2} = \pi^{-1/2}\alpha_{2a1g}^{3/2}$, $\alpha_{2a1g} = Z/2a_0$.

Here $a_0 = \varepsilon\hbar^2/\mu e^2$ is Bohr's radius in a medium of dielectric constant $\varepsilon$, $\mu = 0.5 m_e$ is the electron effective mass.

The mixing constants read:

$u_{ij...k,xyz} = \langle 2t_{1u,xyz} | u_{ij...k}(r) | 1a_{1g}\rangle$

$= A_{\kappa 1}A_{\kappa'0} \int u_{ij...k}(r)\{[sin(\kappa r)/(\kappa r) - cos(\kappa r)]/(\kappa r)\}\{sin(\kappa'r)/(\kappa'r)\}r^2 sin\theta dr d\theta d\varphi^3 |_{r\leq ro}$

$+ B_{21}B_{10} \int u_{ij...k}(r)\mathbf{r}_{xyz}exp[-(\alpha_{2t1u}+\alpha_{1a1g})r]r^2 sin\theta dr d\theta d\varphi |_{r\geq ro}$

$= A_{\kappa 1}A_{\kappa'0} \,_{r0}\!\int^{\infty} dr \,_{0}\!\int^{2\pi} d\varphi \,_{0}\!\int^{\pi} d\theta\, u_{ij...k}(r)\{[sin(\kappa r)/(\kappa r) - cos(\kappa r)]/(\kappa r)\{sin(\kappa'r)/(\kappa'r)\}r^2 sin\theta$

$+ B_{21}B_{10} \,_{r0}\!\int^{\infty} dr \,_{0}\!\int^{2\pi} d\varphi \,_{0}\!\int^{\pi} d\theta\, u_{ij...k}(r)\, \mathbf{r}_{xyz}r^2 sin\theta \, exp[-(\alpha_{2t1u}+\alpha_{1a1g})r]$

$v_{ij...k,xyz} = \langle 2t_{1u,xyz} | u_{ij...k}(r) | 2a_{1g}\rangle$

$= A_{\kappa 1}A_{\kappa'0} \int v_{ij...k}(r)\{[sin(\kappa r)/(\kappa r) - cos(\kappa r)]/(\kappa r)\}\{sin(\kappa'r)/(\kappa'r)\}r^2 sin\theta dr d\theta d\varphi |_{r\leq ro}$

$+ B_{21}B_{20} \int u_{ij...k}(r)\, \mathbf{r}_{xyz}2(1-\alpha_{2a1g}r)exp[-(\alpha_{2t1u}+\alpha_{2a1g})r]r^2 sin\theta dr d\theta d\varphi |_{r\geq ro}$

$= A_{\kappa 1}A_{\kappa'0} \,_{r0}\!\int^{\infty} dr \,_{0}\!\int^{2\pi} d\varphi \,_{0}\!\int^{\pi} d\theta\, u_{ij...k}(r)\{[sin(\kappa r)/(\kappa r) - cos(\kappa r)]/(\kappa r)\{sin(\kappa'r)/(\kappa'r)\}r^2 sin\theta$

$+ B_{21}B_{20} \,_{r0}\!\int^{\infty} dr \,_{0}\!\int^{2\pi} d\varphi \,_{0}\!\int^{\pi} d\theta\, u_{ij...k}(r)\, \mathbf{r}_{xyz}2(1-\alpha_{2a1g}r)r^2 sin\theta \, exp[-(\alpha_{2t1u}+\alpha_{2a1g})r]$

to be derived analytically or calculated numerically. Here $A_{\kappa l}$, $B_{nl}$, the normalization constants, and $\kappa = u / r_0$ are determined by the continuity conditions for the wave functions and their derivatives at $r = r_0$.

## AI.2. Angular considerations

Omitting the normalization factors for the sake of simplicity, we get

$d_{xxx,x} \sim (\alpha_M e^2/r_0^7) \,_{r0}\!\int^{\infty} dr \,_{0}\!\int^{2\pi} d\varphi \,_{0}\!\int^{\pi} d\theta\, x6x(3r_0^2-11x_0^2)(r^2 sin\theta)exp[-(\alpha_{t1u}+\alpha_{a1g})r]$

$= (\alpha_M e^2/r_0^7) \,_{r0}\!\int^{\infty} dr \,_{0}\!\int^{2\pi} d\varphi \,_{0}\!\int^{\pi} d\theta\, 6r^2(cos\varphi sin\theta)^2(r^2 sin\theta)$

$\times r_0^2 [3-11(cos\varphi sin\theta)^2]exp[-(\alpha_{t1u}+\alpha_{a1g})r]$

$d_{xxx,y} \sim (\alpha_M e^2/r_0^7) \,_{r0}\!\int^{\infty} dr \,_{0}\!\int^{2\pi} d\varphi \,_{0}\!\int^{\pi} d\theta\, y6x(3r_0^2-11x_0^2)(r^2 sin\theta)exp[-(\alpha_{t1u}+\alpha_{a1g})r]$

$= (\alpha_M e^2/r_0^7) \,_{r0}\!\int^{\infty} dr \,_{0}\!\int^{2\pi} d\varphi \,_{0}\!\int^{\pi} d\theta\, 6r^2(sin\varphi cos\varphi)(sin\theta)^2(r^2 sin\theta)$

$\times r_0^2[3-11(cos\varphi sin\theta)^2]exp[-(\alpha_{t1u}+\alpha_{a1g})r] = 0 \quad (\varphi)$

$d_{xxx,z} \sim (\alpha_M e^2/r_0^7) \,_{r0}\!\int^{\infty} dr \,_{0}\!\int^{2\pi} d\varphi \,_{0}\!\int^{\pi} d\theta\, z6x(3r_0^2-11x_0^2)(r^2 sin\theta)exp[-(\alpha_{t1u}+\alpha_{a1g})r]$

$= (\alpha_M e^2/r_0^7) \,_{r0}\!\int^{\infty} dr \,_{0}\!\int^{2\pi} d\varphi \,_{0}\!\int^{\pi} d\theta\, 6r^2(cos\varphi sin\theta)(cos\theta)(r^2 sin\theta)$

$\times r_0^2 [3-11(cos\varphi sin\theta)^2]exp[-(\alpha_{t1u}+\alpha_{a1g})r] = 0 \quad (\theta)$

$d_{xxz,x} \sim -(\alpha_M e^2/r_0^7) \,_{r0}\!\int^\infty dr \,_0\!\int^{2\pi} d\varphi \,_0\!\int^\pi d\theta\, x6z_0(rr_0+11xx_0)(r^2 sin\theta)exp[-(\alpha_{t1u}+\alpha_{a1g})r]$

$= -(\alpha_M e^2/r_0^7) \,_{r0}\!\int^\infty dr \,_0\!\int^{2\pi} d\varphi \,_0\!\int^\pi d\theta\, 6rr_0(cos\varphi sin\theta)(cos\theta)(r^2 sin\theta)$

$\times rr_0[1+11(cos\varphi sin\theta)^2]exp[-(\alpha_{t1u}+\alpha_{a1g})r] = 0 \quad (\theta)$

$d_{xxz,y} \sim -(\alpha_M e^2/r_0^7) \,_{r0}\!\int^\infty dr \,_0\!\int^{2\pi} d\varphi \,_0\!\int^\pi d\theta\, y6z_0(rr_0+11xx_0)(r^2 sin\theta)exp[-(\alpha_{t1u}+\alpha_{a1g})r]$

$= -(\alpha_M e^2/r_0^7) \,_{r0}\!\int^\infty dr \,_0\!\int^{2\pi} d\varphi \,_0\!\int^\pi d\theta\, 6rr_0(sin\varphi sin\theta)(cos\theta)(r^2 sin\theta)$

$\times rr_0[1+11(cos\varphi sin\theta)^2]exp[-(\alpha_{t1u}+\alpha_{a1g})r] = 0 \quad (\varphi,\theta)$

$d_{xxz,z} \sim -(\alpha_M e^2/r_0^7) \,_{r0}\!\int^\infty dr \,_0\!\int^{2\pi} d\varphi \,_0\!\int^\pi d\theta\, z6z_0(rr_0+11xx_0)(r^2 sin\theta)exp[-(\alpha_{t1u}+\alpha_{a1g})r]$

$= -(\alpha_M e^2/r_0^7) \,_{r0}\!\int^\infty dr \,_0\!\int^{2\pi} d\varphi \,_0\!\int^\pi d\theta\, 6rr_0(cos\theta)^2(r^2 sin\theta)$

$\times rr_0[1+11(cos\varphi sin\theta)^2]exp[-(\alpha_{t1u}+\alpha_{a1g})r]$

$d_{xyz,x} \sim (\alpha_M e^2/r_0^7) \,_{r0}\!\int^\infty dr \,_0\!\int^{2\pi} d\varphi \,_0\!\int^\pi d\theta\, x74xy_0z_0(r^2 sin\theta)exp[-(\alpha_{t1u}+\alpha_{a1g})r]$

$= (\alpha_M e^2/r_0^7) \,_{r0}\!\int^\infty dr \,_0\!\int^{2\pi} d\varphi \,_0\!\int^\pi d\theta\, 74(rr_0)^2(cos\varphi sin\theta)^2$

$\times (sin\varphi sin\theta)(cos\theta)(r^2 sin\theta)exp[-(\alpha_{t1u}+\alpha_{a1g})r] = 0 \quad (\varphi,\theta)$

$d_{xyz,y} \sim (\alpha_M e^2/r_0^7) \,_{r0}\!\int^\infty dr \,_0\!\int^{2\pi} d\varphi \,_0\!\int^\pi d\theta\, y74xy_0z_0(r^2 sin\theta)exp[-(\alpha_{t1u}+\alpha_{a1g})r]$

$= (\alpha_M e^2/r_0^7) \,_{r0}\!\int^\infty dr \,_0\!\int^{2\pi} d\varphi \,_0\!\int^\pi d\theta\, 74(rr_0)^2(sin\varphi sin\theta)^2$

$\times (cos\varphi sin\theta)(cos\theta)(r^2 sin\theta)exp[-(\alpha_{t1u}+\alpha_{a1g})r] = 0 \quad (\varphi,\theta)$

$d_{xyz,z} \sim (\alpha_M e^2/r_0^7) \,_{r0}\!\int^\infty dr \,_0\!\int^{2\pi} d\varphi \,_0\!\int^\pi d\theta\, z74xy_0z_0(r^2 sin\theta)exp[-(\alpha_{t1u}+\alpha_{a1g})r]$

$= (\alpha_M e^2/r_0^7) \,_{r0}\!\int^\infty dr \,_0\!\int^{2\pi} d\varphi \,_0\!\int \pi d\theta\, 74(rr_0)^2(cos\varphi sin\theta)$

$\times (sin\varphi sin\theta)(cos\theta)^2(r^2 sin\theta)exp[-(\alpha_{t1u}+\alpha_{a1g})r] = 0 \quad (\varphi)$

$c_{xx,x} \sim (\alpha_M e^2/r_0^5) \,_{r0}\!\int^\infty dr \,_0\!\int^{2\pi} d\varphi \,_0\!\int^\pi d\theta\, x3(rr_0+3xx_0)(r^2 sin\theta)exp[-(\alpha_{t1u}+\alpha_{a1g})r]$

$= (\alpha_M e^2/r_0^5) \,_{r0}\!\int^\infty dr \,_0\!\int^{2\pi} d\varphi \,_0\!\int^\pi d\theta\, 3r^2 r_0(cos\varphi sin\theta)(r^2 sin\theta)$

$\times [1+3(cos\varphi sin\theta)^2]exp[-(\alpha_{t1u}+\alpha_{a1g})r] = 0 \quad (\varphi)$

$c_{xx,y} \sim (\alpha_M e^2/r_0^5) \,_{r0}\!\int^\infty dr \,_0\!\int^{2\pi} d\varphi \,_0\!\int^\pi d\theta\, y3(rr_0+3xx_0)(r^2 sin\theta)exp[-(\alpha_{t1u}+\alpha_{a1g})r]$

$= (\alpha_M e^2/r_0^5) \,_{r0}\!\int^\infty dr \,_0\!\int^{2\pi} d\varphi \,_0\!\int^\pi d\theta\, 3r^2 r_0(sin\varphi sin\theta)(r^2 sin\theta)$

$\times [1+3(cos\varphi sin\theta)^2]exp[-(\alpha_{t1u}+\alpha_{a1g})r] = 0 \quad (\varphi)$

$c_{xx,z} \sim (\alpha_M e^2/r_0^5) \int_{r_0}^{\infty} dr \int_0^{2\pi} d\varphi \int_0^{\pi} d\theta\ z3(rr_0+3xx_0)(r^2 sin\theta)exp[-(\alpha_{t1u}+\alpha_{a1g})r]$

$= (\alpha_M e^2/r_0^5) \int_{r_0}^{\infty} dr \int_0^{2\pi} d\varphi \int_0^{\pi} d\theta\ 3r^2 r_0\ (cos\theta)(r^2 sin\theta)$

$\times [1+3(cos\varphi sin\theta)^2]exp[-(\alpha_{t1u}+\alpha_{a1g})r] = 0 \quad (\theta)$

$c_{xy,x} \sim (\alpha_M e^2/r_0^5) \int_{r_0}^{\infty} dr \int_0^{2\pi} d\varphi \int_0^{\pi} d\theta\ x9xy_0(r^2 sin\theta)exp[-(\alpha_{t1u}+\alpha_{a1g})r]$

$= (\alpha_M e^2/r_0^5) \int_{r_0}^{\infty} dr \int_0^{2\pi} d\varphi \int_0^{\pi} d\theta\ 9r^2 r_0\ (cos\varphi sin\theta)^2(sin\varphi sin\theta)$

$\times (r^2 sin\theta)exp[-(\alpha_{t1u}+\alpha_{a1g})r] = 0 \quad (\varphi)$

$c_{xy,y} \sim (\alpha_M e^2/r_0^5) \int_{r_0}^{\infty} dr \int_0^{2\pi} d\varphi \int_0^{\pi} d\theta\ y9xy_0\ (r^2 sin\theta)exp[-(\alpha_{t1u}+\alpha_{a1g})r]$

$= (\alpha_M e^2/r_0^5) \int_{r_0}^{\infty} dr \int_0^{2\pi} d\varphi \int_0^{\pi} d\theta\ 9r^2 r_0\ (cos\varphi sin\theta)(sin\varphi sin\theta)^2$

$\times (r^2 sin\theta)exp[-(\alpha_{t1u}+\alpha_{a1g})r] = 0 \quad (\varphi)$

$c_{xy,z} \sim (\alpha_M e^2/r_0^5) \int_{r_0}^{\infty} dr \int_0^{2\pi} d\varphi \int_0^{\pi} d\theta\ z9xy_0\ (r^2 sin\theta)exp[-(\alpha_{t1u}+\alpha_{a1g})r]$

$= (\alpha_M e^2/r_0^5) \int_{r_0}^{\infty} dr \int_0^{2\pi} d\varphi \int_0^{\pi} d\theta\ 9r^2 r_0\ (cos\varphi sin\theta)(sin\varphi sin\theta)cos\theta$

$\times (r^2 sin\theta)exp[-(\alpha_{t1u}+\alpha_{a1g})r] = 0 \quad (\varphi,\theta)$

$b_{x,x} \sim -(\alpha_M e^2/r_0^3) \int_{r_0}^{\infty} dr \int_0^{2\pi} d\varphi \int_0^{\pi} d\theta\ x2x(r^2 sin\theta)exp[-(\alpha_{t1u}+\alpha_{a1g})r]$

$= -(\alpha_M e^2/r_0^3) \int_{r_0}^{\infty} dr \int_0^{2\pi} d\varphi \int_0^{\pi} d\theta\ 2r^2(cos\varphi sin\theta)^2(r^2 sin\theta)exp[-(\alpha_{t1u}+\alpha_{a1g})r]$

$b_{x,y} \sim -(\alpha_M e^2/r_0^3) \int_{r_0}^{\infty} dr \int_0^{2\pi} d\varphi \int_0^{\pi} d\theta\ y2x(r^2 sin\theta)exp[-(\alpha_{t1u}+\alpha_{a1g})r]$

$= -(\alpha_M e^2/r_0^3) \int_{r_0}^{\infty} dr \int_0^{2\pi} d\varphi \int_0^{\pi} d\theta\ 2r^2(cos\varphi sin\theta)(sin\varphi sin\theta)(r^2 sin\theta)$

$\times exp[-(\alpha_{t1u}+\alpha_{a1g})r] = 0 \quad (\varphi)$

$b_{x,z} \sim -(\alpha_M e^2/r_0^3) \int_{r_0}^{\infty} dr \int_0^{2\pi} d\varphi \int_0^{\pi} d\theta\ z2x(r^2 sin\theta)exp[-(\alpha_{t1u}+\alpha_{a1g})r]$

$= -(\alpha_M e^2/r_0^3) \int_{r_0}^{\infty} dr \int_0^{2\pi} d\varphi \int_0^{\pi} d\theta\ 2r^2(cos\varphi sin\theta)(cos\theta)(r^2 sin\theta)$

$\times exp[-(\alpha_{t1u}+\alpha_{a1g})r] = 0 \quad (\varphi,\theta)$

Angular coordinates in brackets, such as ($\varphi$) or ($\theta$), following an equation for a vanishing coupling constant are those the integration over which makes that constant to vanish.

### AI.3. Coulomb tails for $r \geq r_0$

The integration in r is made using

$\int P_m(r)exp(-ar)dr = -[exp(-ar)/a]\sum_{k=0}^{m} a^{-k}(d^k P_m(r)/dr^k)$,

for $a > 0$ where $P_m(r) = \sum_{i=0}^{m} b_i r_i$ is a polynomial of r, e.g.

$\int r^m exp(-ar)dr = -[exp(-ar)/a][r^m + \sum_{k=1}^{m} a^{-k} m(m-1)...(m-k+1)r^{m-k}]$,

which gives

$_0\int^{\infty} r^m exp(-ar)dr = m!/a^{m+1}$,

$_{r_0}\int^{\infty} dr r^m exp(-ar) = [exp(-ar_0)/a^{m+1}][(ar_0)^m + \sum_{k=1}^{m} m(m-1)...(m-k)(ar_0)^{m-k}]$.

Introducing $N_{a1g}N_{t1u} = (1/4\pi\sqrt{2})(Z/a_0)^4$ we derive the nonvanishing mixing constants as follows:

$b_{x,x} \sim -N_{a1g}N_{t1u}(\alpha_M e^2/r_0) \, _{r_0}\!\int^{\varphi} dr \, _0\!\int^{2\pi} d\varphi \, _0\!\int^{\pi} d\theta \, 2r^2(cos\varphi sin\theta)^2 (r^2 sin\theta)$

$\times exp[-(\alpha_{t1u}+\alpha_{a1g})r]$

$= -2N_{a1g}N_{t1u}(\alpha_M e^2/r_0^3) \, _0\!\int^{2\pi} d\varphi (cos\varphi)^2 \, _0\!\int^{\pi} d\theta (sin\theta)^3$

$\times [exp(-ar_0)/a^5][(ar_0)^4 + \sum_{k=1}^{4} 4\times 3\times...\times(4-k+1)(ar_0)^{4-k}]$

$= -N_{a1g}N_{t1u}(\alpha_M e^2/r_0^3)(8\pi/3)[exp(-ar_0)/a^5][(ar_0)^4 + \sum_{k=1}^{4} 4\times 3\times...\times(4-k+1)(ar_0)^{4-k}]$

$= -(8\pi/3)(V_M/r_0)N_{a1g}N_{t1u}(1/a^4)[exp(-ar_0)/(ar_0)][(ar_0)^4 + \sum_{k=1}^{4} 4\times 3\times...\times(4-k+1)(ar_0)^{4-k}]$,

$d_{xxx,x} \sim N_{a1g}N_{t1u}(\alpha_M e^2/r_0^7) \, _{r_0}\!\int^{\infty} dr \, _0\!\int 2\pi d\varphi \, _0\!\int \pi d\theta \, 6r^2 (cos\varphi sin\theta)^2 (r^2 sin\theta)$

$\times r_0^2 [3-11(cos\varphi sin\theta)^2]exp[-(\alpha_{t1u}+\alpha_{a1g})r]$

$= 6N_{a1g}N_{t1u}(\alpha_M e^2/r_0^5) \, _0\!\int^{2\pi} d\varphi (cos\varphi)^2 \, _0\!\int^{\pi} d\theta (sin\theta)^3 [3-11(cos\varphi sin\theta)^2]$

$\times [exp(-ar_0)/a^5][(ar_0)^4 + \sum_{k=1}^{4} 4\times 3\times...\times(4-k+1)(ar_0)^{4-k}]$

$= 6N_{a1g}N_{t1u}(\alpha_M e^2/r_0^5)[exp(-ar_0)/a^5][(ar_0)^4 + \sum_{k=1}^{4} 4\times 3\times...\times(4-k+1)(ar_0)^{4-k}]$

$\times \, _0\!\int^{2\pi} d\varphi \, _0\!\int^{\pi} d\theta [3(cos\varphi)^2(sin\theta)^3 - 11(cos\varphi)^4(sin\theta)^5]$

$= 6N_{a1g}N_{t1u}(\alpha_M e^2/r_0^5)[exp(-ar_0)/a^5][(ar_0)^4 + \sum_{k=1}^{4} 4\times 3\times...\times(4-k+1)(ar_0)^{4-k}]$

$\times [3\pi(-2)(-2/3) - 11(3/8)\pi(-2)((4/15)-(4/5))]$

$= -(12/5)\pi N_{a1g}N_{t1u}(\alpha_M e^2/r_0^5)[exp(-ar_0)/a^5][(ar_0)^4 + \sum_{k=1}^{4} 4\times 3\times...\times(4-k+1)(ar_0)^{4-k}]$

$= -N_{a1g}N_{t1u}(12/5)\pi(V_M/r_0^3)(1/a^4)[exp(-ar_0)/(ar_0)][(ar_0)^4 + \sum_{k=1}^{4} 4\times 3\times...\times(4-k+1)(ar_0)^{4-k}]$

$d_{xxz,z} \sim -N_{a1g}N_{t1u}(\alpha_M e^2/r_0^7) \, _{r_0}\!\int^{\infty} dr \, _0\!\int^{2\pi} d\varphi \, _0\!\int^{\pi} d\theta \, 6rr_0 (cos\theta)^2 (r^2 sin\theta)$

$\times rr_0 [1+11(cos\varphi sin\theta)^2]exp[-(\alpha_{t1u}+\alpha_{a1g})r]$

$= -6N_{a1g}N_{t1u}(\alpha_M e^2/r_0^5) \int_0^{2\pi} d\varphi \int_0^{\pi} d\theta(cos\theta)^2(sin\theta)[1+11(cos\varphi sin\theta)^2]$

$\times [exp(-ar_0)/a^5][(ar_0)^4+\sum_{k=1}^{4} 4\times 3\times...\times(4-k+1)(ar_0)^{4-k}]$

$= -6N_{a1g}N_{t1u}(\alpha_M e^2/r_0^5) \int_0^{2\pi} d\varphi \int_0^{\pi} d\theta[(cos\theta)^2(sin\theta)+11(cos\varphi)^2(cos\theta)^2(sin\theta)^3]$

$\times [exp(-ar_0)/a^5][(ar_0)^4+\sum_{k=1}^{4} 4\times 3\times...\times(4-k+1)(ar_0)^{4-k}]$

$= -N_{a1g}N_{t1u} 6\pi(64/15)[\alpha_M e^2/r_0 5]\times [exp(-ar_0)/a^5][(ar_0)^4+\sum_{k=1}^{4} 4\times 3\times...\times(4-k+1)(ar_0)^{4-k}]$

$= -N_{a1g}N_{t1u} 6\pi(64/15)(V_M/r_0^3)\times(1/a^4) [exp(-ar_0)/(ar_0)]$

$\times [(ar_0)^4+\sum_{k=1}^{4} 4\times 3\times...\times(4-k+1)(ar_0)^{4-k}].$

Setting $\alpha=\alpha_{2t1u}+\alpha_{1a1g}= (3/2)(Z/a_0)$ or $\alpha=\alpha_{2t1u}+\alpha_{2a1g} = (Z/a_0)$, we get $N_{a1g}N_{t1u}\times (1/a^4) = (2/3)^4 (1/4\pi\sqrt{2})$ for 1s-2p and $N_{a1g}N_{t1u}(1/a^4) = (1/4\pi\sqrt{2})$ for 2s-2p.

### AI.4. In-cavity contribution for $r \leq r_0$

The following radial dependence is generated by the spherical Bessel functions:

$\{[sin(\kappa r)/(\kappa r)-cos(\kappa r)]/(\kappa r)\}[sin(\kappa' r)/(\kappa' r)] =$

$sin(\kappa r)sin(\kappa' r)/(\kappa r)^2(\kappa' r)-cos(\kappa r)sin(\kappa' r)/(\kappa r)(\kappa' r) =$

$½\{[(\kappa r)^2(\kappa' r)]^{-1}(cos[½(\kappa-\kappa')r]-cos[½(\kappa+\kappa')r]) -$

$[(\kappa r)(\kappa' r)]^{-1}(sin[½(\kappa-\kappa')r]+sin[½(\kappa+\kappa')r])\}$

which results in the radial integral:

$P(v,v') = A_{\kappa 1}A_{\kappa'0} \int_0^{r_0} dr\ ½\{(\kappa^2\kappa')^{-1}(cos[½(\kappa-\kappa')r]-cos[½(\kappa+\kappa')r]) -$

$(\kappa\kappa')^{-1}r(sin[½(\kappa-\kappa')r]+sin[½(\kappa+\kappa')r])\} =$

$A_{\kappa 1}A_{\kappa'0} ½(\kappa^2\kappa')^{-1}\{[½(\kappa-\kappa')]^{-1}sin[½(\kappa-\kappa')r_0] -$

$[½(\kappa+\kappa')]^{-1}sin[½(\kappa+\kappa')r_0]\} + A_{\kappa 1}A_{\kappa'0} ½(\kappa\kappa')^{-1}\{[½(\kappa-\kappa')]^{-1}(r_0 cos[½(\kappa-\kappa')r_0]$

$- [½(\kappa-\kappa')]^{-1}sin[½(\kappa-\kappa')r_0]) + [½(\kappa+\kappa')]^{-1}(r_0 cos[½(\kappa+\kappa')r_0] -$

$[½(\kappa+\kappa')]^{-1}sin[½(\kappa+\kappa')r_0])\} =$

$A_{\kappa 1}A_{\kappa'0} ½r_0^3 (v^2v')^{-1}\{[½(v-v')/r_0]^{-1}sin[½(v-v')] - [½(v+v')/r_0]^{-1}sin[½(v+v')]\} +$

$$A_{\kappa 1}A_{\kappa' 0}\tfrac{1}{2}r_0^2(vv')^{-1}\{[\tfrac{1}{2}(v-v')/r_0]^{-1}(r_0 cos[\tfrac{1}{2}(v-v')] -$$

$$[\tfrac{1}{2}(v-v')/r_0]^{-1}sin[\tfrac{1}{2}(v-v')]) + [\tfrac{1}{2}(v+v')/r_0]^{-1}(r_0 cos[\tfrac{1}{2}(v+v')] - [\tfrac{1}{2}(v+v')/r_0]^{-1}sin[\tfrac{1}{2}(v+v')])\}$$

$$= A_{\kappa 1}A_{\kappa' 0} \tfrac{1}{2}r_0^4 \{(v^2v')^{-1}\{[\tfrac{1}{2}(v-v')]^{-1}sin[\tfrac{1}{2}(v-v')] - [\tfrac{1}{2}(v+v')]^{-1}sin[\tfrac{1}{2}(v+v')]\}$$

$$+ (vv')^{-1}\{[\tfrac{1}{2}(v-v')]^{-1}(cos[\tfrac{1}{2}(v-v')] - [\tfrac{1}{2}(v-v')]^{-1}sin[\tfrac{1}{2}(v-v')]) + [\tfrac{1}{2}(v+v')]^{-1}(cos[\tfrac{1}{2}(v+v')]$$

$$- [\tfrac{1}{2}(v+v')]^{-1}sin[\tfrac{1}{2}(v+v')])\}\}.$$

Note that $P(v,v') \propto r_0$ in so far as $A_{\kappa 1}A_{\kappa' 0} \propto r_0^{-3}$. Using $P(v,v')$ and accounting for the angular dependences, we get for $r \leq r_0$:

$$b_{x,x} \sim -(\alpha_M e^2/r_0^3) \int_0^{2\pi} d\varphi \int_0^{\pi} d\theta\, 2(cos\varphi sin\theta)^2 sin\theta\, P(v,v')$$

$$= -(8\pi/3)(\alpha_M e^2/r_0^3)P(v,v') = -(8\pi/3)(V_M/r_0^2)P(v,v');$$

$$d_{xxx,x} \sim (\alpha_M e^2/r_0^7) \int_0^{2\pi} d\varphi \int_0^{\pi} d\theta\, 6(cos\varphi sin\theta)^2 sin\theta\, r_0^2[3-11(cos\varphi sin\theta)^2]\, P(v,v')$$

$$= -(24\pi/5)(\alpha_M e^2/r_0^5)P(v,v') = -(24\pi/5)(V_M/r_0^4)P(v,v');$$

$$d_{xxz,z} \sim -(\alpha_M e^2/r_0^7) \int_0^{2\pi} d\varphi \int_0^{\pi} d\theta\, 6r_0 (cos\theta)^2 sin\theta\, r_0 [1+11(cos\varphi sin\theta)^2]\, P(v,v')$$

$$= -4(1+22\pi/5)(\alpha_M e^2/r_0^5)P(v,v') = -4(1+22\pi/5)(V_M/r_0^4)P(v,v').$$

### AI.5. Complete mixing constants

Values for the semicontinuum normalization constants $A_{Kl}$ and $B_{nl}$ satisfying the continuity conditions at $r=r_0$ are given in Table I following Ref. 10. Comparing the notations used therein and presently, we have the following relationship with the all-space normalization constants ($u = \alpha r_0$):

$$N_{1a1g} = \pi^{-1/2}(Z/a_0)^{3/2} = \pi^{-1/2}\alpha_{1a1g}^{3/2} = (1/\sqrt{\pi})u_{1a1g}^{3/2}r_0^{3/2} = (4\pi)^{-1/2}B_{10}$$

$$N_{2t1u} = \pi^{-1/2}(Z/2a_0)^{5/2} = \pi^{-1/2}\alpha_{2t1u}^{5/2} = (1/\sqrt{\pi})\alpha_{2t1u}u_{2t1u}^{3/2}r_0^{-3/2} = (3/4\pi)^{1/2}\alpha_{2t1u}B_{21}$$

$$N_{2a1g} = \pi^{-1/2}(Z/2a_0)^{3/2} = \pi^{-1/2}\alpha_{2a1g}^{3/2} = (1/\sqrt{\pi})u_{2a1g}^{3/2}r_0^{-3/2} = (4\pi)^{-1/2}B_{20}$$

The complete mixing constants deriving from the arguments in AI.3 and AI.4 are calculated as given in Section 4 of the basic text.

### AI.6. Second-order mixing of same-parity states (pseudo-Renner effect)

States of the same parity may only mix through the even-order coupling terms in the interaction expansion. Accordingly, we define second-order mixing constants between same-parity states, even-even and odd-odd, viz.:

$$c_{1s2s} = <1s | \partial^2 V(r_0,q_l)/\partial q_x^2 |_{q=0} | 2s> \equiv <1s | 3(rr_0+3xx_0)/r_0^5 | 2s>$$

$c_{xy} = \langle 2p_x | \partial^2 V(r_0,q_l)/\partial q_x \partial q_y |_{q=0} | 2p_y \rangle = \langle 2p_x | 9xy_0/r_0^5 | 2p_y \rangle$,

etc. These will be computed using the semicontinuum wavefunctions as above. For the in-cavity contribution, we have ($\alpha_M e^2$ dropped)

$c_{1s2s}(r \leq r_0) = (3/r_0^4)A_{\kappa 0}A_{\kappa' 0} \int_0^{r_0} dr\, r^3 \int_0^{2\pi} d\varphi \int_0^{\pi} d\theta\, [1+3(cos\varphi sin\theta)^2]sin\theta$

$\times [sin(\kappa r)/(\kappa r)][sin(\kappa' r)/(\kappa' r)]$

$= 3A_{\kappa 0}A_{\kappa' 0} \int_0^1 dp\, p^3 \int_0^{2\pi} d\varphi \int_0^{\pi} d\theta\, [1+3(cos\varphi sin\theta)^2]sin\theta$

$\times [sin(vp)/(vp)][sin(v'p)/(v'p)]$

$= 24\pi \int_0^1 dp\, p^3\, A_{\kappa 0}A_{\kappa' 0} [sin(vp)/(vp)][sin(v'p)/(v'p)]$

$= (12\pi/vv')A_{\kappa 0}A_{\kappa' 0} \int_0^1 dp\, p[cos((v-v')p) - cos((v+v')p)]$

$= (12\pi/vv')A_{\kappa 0}A_{\kappa' 0} \{[(v-v')^{-1}sin(v-v') - (v+v')^{-1}sin(v+v')]$

$+ [(v-v')^{-2}cos(v-v') - (v+v')^{-2}cos(v+v')]\}$

and we similarly derive for the out-of-cavity part:

$c_{1s2s}(r \geq r_0) = 24\pi^3 \int_1^{\infty} dp\, p^3\, B_{10}B_{20}\, exp(-u_{1s}p)2(1-u_{2s}p)exp(-u_{2s}p)$

$= 48\pi B_{10}B_{20}r_0^{-4} \int_{r_0}^{\infty} dr\, r^3 (1-\alpha_{2s}r)exp(-[\alpha_{1s}+\alpha_{2s}]r)$

$= 48\pi B_{10}B_{20} [\mathbf{exp}(-[u_{1s}+u_{2s}])/(u_{1s}+u_{2s})^4]\{[u^3+\sum_{k=1}^{3} 3\times 2\times ...\times(3-k)u^{3-k}]$

$- [u_{2s}/(u_{1s}+u_{2s})][u^4+\sum_{k=1}^{4} 4\times 3\times ...\times(4-k)u^{4-k}]\}$

with $\alpha = \alpha_{1s}+\alpha_{2s}$, $u = \alpha r_0 = u_{1s}+u_{2s}$. Combining we get

$c_{1s2s} = \alpha_M e^2\{(12\pi/vv')A_{\kappa 0}A_{\kappa' 0}\{[(v-v')^{-1}sin(v-v') - (v+v')^{-1}sin(v+v')]$

$+ [(v-v')^{-2}cos(v-v') - (v+v')^{-2}cos(v+v')]\} + B_{10}B_{20}48\pi[exp(-[u_{1s}+u_{2s}])/(u_{1s}+u_{2s})^4]$

$\times \{[u^3+\sum_{k=1}^{3} 3\times 2\times ...\times(3-k)u^{3-k}] - [u_{2s}/(u_{1s}+u_{2s})][u^4+\sum_{k=1}^{4} 4\times 3\times ...\times(4-k)u^{4-k}]\}\}$.

with $\alpha = \alpha_{1s}+\alpha_{2s}$, $u = \alpha r_0 = u_{1s}+u_{2s}$.

We similarly derive the other mixing constant. Out of cavity:

$c_{xy}(r \geq r_0) = \int_{r_0}^{\infty} dr \int_0^{2\pi} d\varphi \int_0^{\pi} d\theta\, (x_0/r_0)(y_0/r_0)[9xy_0/r_0^5]r^2 sin\theta\, B_{21}B_{21}exp(-2\alpha_{2p}r)$

$= (9/r_0^4) \int_{r_0}^{\infty} dr \int_0^{2\pi} d\varphi \int_0^{\pi} d\theta\, r^3 (cos\varphi sin\theta)^2(sin\varphi sin\theta)^2 sin\theta\, B_{21}B_{21}exp(-2\alpha_{2p}r)$

$$= (12/5r_0^4)\pi B_{21}B_{21}\,_{r_0}\!\int^{\infty} dr r^3 exp(-2\alpha_{2p}r)$$

$$= (12/5)\pi B_{21}B_{21}\,[exp(-u)/u^4][u^3+\sum_{k=1}^{3} 3\times 2\times...\times(3-k)u^{3-k}]$$

with $\alpha = 2\alpha_{2p}$, $u = \alpha r_0 = 2\alpha_{2p}r_0 = 2u_{2p}$. In-cavity:

$$c_{xy}(r\le r_0) = \,_3{\!\int^{r_0}} dr \,_0{\!\int^{2\pi}} d\varphi \,_0{\!\int^{\pi}} d\theta\, r^2 sin\theta\, (x_0/r_0)(y_0/r_0)[9xy_0/r_0^5]$$

$$\times A_{\kappa 1}A_{\kappa' 1}\,[sin(\kappa r)/(\kappa r)-cos(\kappa r)][sin(\kappa' r)/(\kappa' r)-cos(\kappa' r)]/(\kappa r)(\kappa' r)$$

$$= (9/r_0^4)^3 \,_0{\!\int^{r_0}} dr \,_0{\!\int^{2\pi}} d\varphi \,_0{\!\int^{\pi}} d\theta\, r^3 sin\theta (cos\varphi sin\theta)^2(sin\varphi sin\theta)^2$$

$$\times A_{\kappa 1}A_{\kappa' 1}\,[sin(\kappa r)/(\kappa r)sin(\kappa' r)/(\kappa' r)-sin(\kappa r)/(\kappa r)cos(\kappa' r)$$

$$- cos(\kappa r)sin(\kappa' r)/(\kappa' r)+cos(\kappa r)cos(\kappa' r)]$$

$$= (6\pi/5r_0^4)A_{\kappa 1}A_{\kappa' 1}\,_0{\!\int^{r_0}} dr\{r[cos(\kappa-\kappa')r-cos(\kappa+\kappa')r]/\kappa\kappa'$$

$$- r^2\,[sin(\kappa+\kappa')r+sin(\kappa-\kappa')r]/\kappa - r^2\,[sin(\kappa+\kappa')r+sin(\kappa-\kappa')r]/\kappa'$$

$$+ r^3\,[cos(\kappa+\kappa')r+cos(\kappa-\kappa')r]\}$$

$$= (6\pi/5r_0^4)A_{\kappa 1}A_{\kappa' 1}\{\{r_0\,[(\kappa-\kappa')^{-1}sin(\kappa-\kappa')r_0 - (\kappa+\kappa')^{-1}sin(\kappa+\kappa')r_0]$$

$$+ (\kappa-\kappa')^{-2}sin(\kappa-\kappa')r_0-(\kappa+\kappa')^{-2}sin(\kappa+\kappa')r_0]\}/\kappa\kappa'$$

$$+ (1/\kappa+1/\kappa')\sum_{k=0}^{2} k!(2\ k)r_0^{2-k}\{(\kappa+\kappa')^{-(k+1)}cos([\kappa+\kappa']r_0+½k\pi)$$

$$+ (\kappa-\kappa')^{-(k+1)}cos([\kappa-\kappa']r_0+½k\pi)\}$$

$$+ \sum_{k=0}^{3} k!(3\ k)r_0^{3-k}\{(\kappa+\kappa')^{-(k+1)}sin([\kappa+\kappa']r_0+½k\pi) + (\kappa-\kappa')^{-(k+1)}sin([\kappa-\kappa']r_0+½k\pi)\}\}$$

$$= (6\pi/5)A_{\kappa 1}A_{\kappa' 1}\{\{[(v-v')^{-1}sin(v-v')-(v+v')^{-1}sin(v+v')]$$

$$+ [(v-v')^{-2}sin(v-v')-(v+v')^{-2}sin(v+v')]\}/vv'$$

$$+ (1/v+1/v')\sum_{k=0}^{2} k!(2\ k)\{(v+v')^{-(k+1)}cos(v+v'+½k\pi) + (v-v')^{-(k+1)}cos(v-v'+½k\pi)\}$$

$$+ \sum_{k=0}^{3} k!(3\ k)\{(v+v')^{-(k+1)}sin(v+v'+½k\pi)+(v-v')^{-(k+1)}sin(v-v'+½k\pi)\}\}.$$

At $v=v'$ ($\kappa=\kappa'$) we have instead

$$c_{xy}(r\le r_0) = (6\pi/5r_0^4)A_{\kappa 1}A_{\kappa' 1}\,_0{\!\int^{r_0}} dr\{r[1-cos(\kappa+\kappa')r]/\kappa\kappa'$$

$$- r^2\,[sin(\kappa+\kappa')r]/\kappa - r^2\,[sin(\kappa+\kappa')r]/\kappa' + r^3\,[1+cos(\kappa+\kappa')r]\}$$

$$= (6\pi/5r_0^4)A_{\kappa 1}A_{\kappa' 1}\{½r_0^2/\kappa^2 - [(2\kappa)^{-1}r_0 sin(2\kappa r_0)+(2\kappa)^{-2}cos(2\kappa r_0)]/\kappa^2$$

$- (2/\kappa)\sum_{k=0}^{2} k!(2\ k)r_0^{2-k} (2\kappa)^{-(k+1)} cos(2\kappa r_0 + \tfrac{1}{2}k\pi)$

$+ \sum_{k=0}^{3} k!(3\ k)r_0^{3-k} (2\kappa)^{-(k+1)} sin(2\kappa r_0 + \tfrac{1}{2}k\pi) \}$

$= (6\pi/5)A_{\kappa 1}A_{\kappa' 1} \times \{ \tfrac{1}{2}/v^2 - [(2v)^{-1}sin(2v)+(2v)^{-2}cos(2v)]/v^2$

$- (2/v)\sum_{k=0}^{2} k!(2\ k)(2v)^{-(k+1)} cos(2v+\tfrac{1}{2}k\pi) + \sum_{k=0}^{3} k!(3\ k)(2v)^{-(k+1)} sin(2v+\tfrac{1}{2}k\pi) \}$

Combining we obtain at v≠v':

$c_{xy} = \alpha_M e^2 \{ (6\pi/5)A_{\kappa 1}A_{\kappa' 1} \{ \{ [(v-v')^{-1}sin(v-v') - (v+v')^{-1}sin(v+v')]$

$+ [(v-v')^{-2}sin(v-v') - (v+v')^{-2}sin(v+v')] \}/vv'$

$+ (1/v+1/v')\sum_{k=0}^{2} k!(2\ k)\{(v+v')^{-(k+1)}cos(v+v'+\tfrac{1}{2}k\pi)$

$+ (v-v')^{-(k+1)}cos(v-v'+\tfrac{1}{2}k\pi)\} + \sum_{k=0}^{3} k!(3\ k)\{(v+v')^{-(k+1)}sin(v+v'+\tfrac{1}{2}k\pi)$

$+ (v-v')^{-(k+1)}sin(v-v'+\tfrac{1}{2}k\pi) \} \}$

$+ (12/5)\pi B_{21}B_{21} [exp(-u)/u^4][u^3+\sum_{k=1}^{3} 3\times2\times...\times(3-k)u^{3-k}] \}$

while at v=v':

$c_{xy} = \alpha_M e^2 \{ (6\pi/5)A_{\kappa 1}A_{\kappa' 1} \times \{ \tfrac{1}{2}/v^2 - [(2v)^{-1}sin(2v)+(2v)^{-2}cos(2v)]/v^2$

$- (2/v)\sum_{k=0}^{2} k!(2\ k)(2v)^{-(k+1)}cos(2v+\tfrac{1}{2}k\pi) + \sum_{k=0}^{3} k!(3\ k)(2v)^{-(k+1)}sin(2v+\tfrac{1}{2}k\pi) \}$

$+ (12/5)\pi B_{21}B_{21} [exp(-u)/u^4][u^3+\sum_{k=1}^{3} 3\times2\times...\times(3-k)u^{3-k}] \}$

with $\alpha = 2\alpha_{2p}$, $u = \alpha r_0 = 2\alpha_{2p}r_0 = 2u_{2p}$. There we have constants for pseudo-Renner-effect mixing of the first three semi-continuum eigenstates of the F center in alkali halides.